\documentstyle[11pt]{article}
\def\appendix{\renewcommand{\thesection}{\Alph{section}}\setcounter{section}{0}
              \renewcommand{\theequation}
            {\mbox{\Alph{section}.\arabic{equation}}}\setcounter{equation}{0}}
\def\maketitle{\thispagestyle{empty}\setcounter{page}0\newpage
                \renewcommand{\thefootnote}{\arabic{footnote}}
                  \setcounter{footnote}0}
\renewcommand{\thanks}[1]{\renewcommand{\thefootnote}{\fnsymbol{footnote}}
               \footnote{#1}\renewcommand{\thefootnote}{\arabic{footnote}}}
\newcommand{\preprint}[1]{\hfill{\sl preprint - #1}\par\bigskip\par\rm}
\renewcommand{\title}[1]{\begin{center}\Large\bf #1\end{center}\rm\par\bigskip}
\renewcommand{\author}[1]{\begin{center}\Large #1\end{center}}
\newcommand{\address}[1]{\begin{center}\large #1\end{center}}

\def\dinfn{\smallskip $^3$ Dipartimento di Fisica, Universit\`a di Trento\\ 
                           and Istituto Nazionale di Fisica Nucleare,\\
                                   Gruppo Collegato di Trento, Italia}

\def\Idinfn{\address{\dinfn}}
\def\dinbcn{\smallskip $^1$ Consejo Superior de Investigaciones
 Cient\'{\i}ficas \\
 IEEC, Edifici Nexus 201, Gran Capit\`a 2-4, 08034 Barcelona, Spain\\
 and Departament ECM and IFAE, Facultat de F\'{\i}sica, \\
 Universitat de Barcelona, Diagonal 647, 08028 Barcelona, Spain}
\def\Idinbcn{\address{\dinbcn}}
\newcommand{\email}[1]{e-mail: \sl #1@science.unitn.it\rm}
\newcommand{\zmail}[1]{e-mail: \sl #1@ieec.fcr.es\rm}
\newcommand{\fzmail}[1]{\thanks{\zmail{#1}}}
\newcommand{\femail}[1]{\thanks{\email{#1}}}
\newcommand{\pacs}[1]{\smallskip\noindent{\sl PACS numbers:
                       \hspace{0.3cm}#1}\par\bigskip\rm}
\def\babs{\hrule\par\begin{description}\item{Abstract: }\it}
\def\eabs{\par\end{description}\hrule\par\medskip\rm}
\renewcommand{\date}[1]{\par\bigskip\par\sl\hfill #1\par\medskip\par\rm}
\newcommand{\ack}[1]{\par\section*{Acknowledgements} #1}
\def\Idinbcn{\address{\dinbcn}}
\def\dimpcoll{\smallskip $^2$ Theoretical Physics Group, Imperial College, \\
Prince Consort Road, London SW7 2BZ, U.K.} 
\def\Idimpcoll{\address{\dimpcoll}}

\newcommand{\icmail}[1]{e-mail: \sl #1@ic.ac.uk\rm}
\newcommand{\ficmail}[1]{\thanks{\icmail{#1}}}
\def\eabs{\par\end{description}\hrule\par\medskip\rm}
\renewcommand{\date}[1]{\par\bigskip\par\sl\hfill #1\par\medskip\par\rm}

\def\nn{\nonumber}            
\def\beq{\begin{eqnarray}}    
\def\eeq{\end{eqnarray}}      
\def\at{\left(}               
\def\aq{\left[}               
\def\ct{\right)}              
\def\cq{\right]}              
\def\R{{\hbox{{\rm I}\kern-.2em\hbox{\rm R}}}}   
\def\H{{\hbox{{\rm I}\kern-.2em\hbox{\rm H}}}}   
\def\N{{\hbox{{\rm I}\kern-.2em\hbox{\rm N}}}}   
\def\C{{\ \hbox{{\rm I}\kern-.6em\hbox{\bf C}}}} 
\def\Z{{\hbox{{\rm Z}\kern-.4em\hbox{\rm Z}}}}   
\def\ii{\infty}                                  
\def\Tr{\mathop{\rm Tr}\nolimits}                  
\renewcommand{\Re}{\mathop{\rm Re}\nolimits}       
\def\lap{\Delta}                                   

\def\ep{\varepsilon}
\def\ze{\zeta}

\def\Ga{\Gamma}

\begin{document}

\preprint{Imperial/TP/97-98/37 }

\title{Is the multiplicative anomaly dependent on the regularization~?}

\author{Emilio Elizalde$^{1,}$\fzmail{elizalde}\,, \\
Antonio Filippi$^{2,}$\ficmail{a.filippi} \,, \\
Luciano Vanzo$^{3,}$\femail{vanzo} and
Sergio Zerbini$^{3,}$\femail{zerbini}
}
\Idinbcn
\Idimpcoll
\Idinfn

\date{April 1998}

\babs
In a recent work, T.S. Evans has claimed that the multiplicative anomaly 
associated with the zeta-function regularization of functional 
determinants is regularization dependent. We show that, if one makes 
use of  consistent definitions, this is not 
the case and clarify  some points in Evans' argument.
\eabs

\pacs{05.30.-d,05.30.Jp,11.10.Wx,11.15.Ex}

Recently, T.S. Evans \cite{evan98} has investigated the role of
 the multiplicative 
anomaly in quantum field theory. His conclusions may be summarized by saying 
that the multiplicative anomaly is regularization dependent and can have 
therefore no physical relevance. 

In our opinion, as far as the zeta-function regularization issue is 
concerned, the starting point of Evans' considerations is not the 
more appropriate , since he assumes that  the (Euclidean) 
one-loop partition function (generating functional) is (formally) given by
\beq
Z= \exp{\at -\frac{1}{2} \Tr \ln A \ct}
\:,\label{e1}\eeq
where $A$ is some elliptic operator (the small disturbances operator). 
However, it is well known that formally
\beq
Z= \at \det A \ct^{-1/2}
\:,\label{e2}\eeq 
The quantity $ \det A$ is ill defined. Zeta-function regularization 
was introduced in order to give a 
proper mathematical meaning to it. As is  known, it consists in 
analytically continuing the quantity 
$\ze(s|A)=\Tr A^{-s}$, perfectly meaningful for $\Re s$ sufficiently large 
and for $A$ elliptic,
and in assuming then, by 
definition, that  \cite{dowk76-13-3224,hawk77-55-133} 
\beq
\ln \det A=-\ze'(0|A)
\:.\label{e4}\eeq
This is possible because, at $s=0$,  the analytical continuation of $\Tr A^{-s}$  is 
regular. From now on, for the sake of clarity, we will call zeta-function 
regularization the above well established mathematical procedure.

One might argue that the starting point ~(\ref{e2}) is  also not 
well defined, being 
based on a formal divergent quantity and the starting point given by 
Eq.~(\ref{e1}) is also acceptable. 
However, the identity $\Tr \ln A= \ln \det A$, valid for a 
positive hermitian matrix, does not necessarily hold in the infinite dimensional case. 
In fact, both  quantities are ill defined, the second one can be 
successfully treated by the zeta-function regularization method. What about 
the first? With regard to this issue, it is easy to see that one of the  
simplest  analytical 
regularizations of the ill defined quantity $ \Tr \ln A$, 
{\it preserving} the linear property of the trace, which is necessary 
for a reasonable definition of a regularized trace, may be the following
\beq
T(A)(s)= \Tr (\ln(A) Q^{-s})
\:,\label{e5}\eeq
where $Q$ is some elliptic operator, acting as a regulator. However, the 
analytical continuation of $T(A)(s)$, since $\ln A$ is a 
pseudo-differential operator, has always a simple pole at $s=0$, and 
the regularization parameter cannot be safely removed. This  also shows that the formal identity
\beq
\Tr \ln A= \ln \det A
\:,\label{e7}\eeq 
cannot have  general validity. As an 
example in $\R^4$ (strictly speaking one should work only
in a compact manifold, but working in $\R^D$ just amounts to factorizing 
a trivial infinite volume, which we shall here safely ignore), 
take $A=L+V$, $L=-\lap$, $V$ constant, 
$Q=-\lap$. Thus the spectral theorem gives
\beq
T(A)(s) \equiv \int_0^\ii dk\,  k^3 \ln (k^2+V)\, k^{-2s} \equiv 
\frac{\Ga(2-s)\Ga(2s-1)}{2(2-2s)}V^{2-2s}
\:.\label{e6}\eeq 
This quantity has a pole at $s=0$. 
Let us  continue with  
Evans' starting point, namely  Eq.~(\ref{e1}), bearing in mind that, 
since the mathematics is different, also the physical consequences 
might be different. However, since it is potentially dangerous 
to perform  manipulations on 
divergent quantities, let us repeat his argument making now use of {\it 
regularized} 
quantities, in the spirit of the analytic regularization method.

Let us consider $A=L+V$ and $B=L$. This is a simpler case than the one 
treated by Evans, but the conclusions are just the same. We have to define a 
regularized  multiplicative anomaly, $b(s)$, as
\beq
b(s)=\frac{\partial T(AB)}{\partial V}-\frac{\partial T(A)}{\partial 
V}-\frac{\partial T(B)}{\partial V}
\:,\label{e8}\eeq
and eventually try to remove the cutoff parameter $s \to 0$ after the 
analytical continuation has been performed.
For suitable $\Re s$, all the integrals exist, and a simple calculation gives 
\beq
b(s)&\equiv& \int_0^\ii dk k^3 \frac{k^2}{(k^2+V)k^2} k^{-2s}- 
\int_0^\ii dk k^3 \frac{1}{(k^2+V)} k^{-2s} \nn \\
&=&\frac{1}{2}\Ga(1-s)\Ga(s)V^{-s}-\frac{1}{2}\Ga(1-s)\Ga(s)V^{-s}=0
\:.\label{e9}\eeq
As a result, the $b$ anomaly introduced by Evans is 
vanishing. This result is obviously  consistent with the linear 
property of the trace and with the operator identity $\ln AB=\ln A+\ln B$, 
valid if $A$ and $B$ commute.

On the other hand, it is easy to see that if,
 after some formal manipulation (taking the derivative of a divergent 
quantity ), one introduces 
in the formal definition of $b$ an analytical regularization which 
{\it does not preserve} the 
linearity of the trace (e.g., changes  the analytic structure of the 
propagator) then  
one can certainly obtain a non vanishing $b$ anomaly, as Evans has shown 
\cite{evan98}. By the way, strictly 
speaking, the analytical regularization used by Evans is 
not the zeta-function regularization method.

Concerning  the multiplicative property of determinants, 
the zeta-function regularized 
determinant $\ln \det A$ suffers from the presence of a 
multiplicative anomaly 
(see, for example, \cite{kass89-177-199,eliz97u-394}), namely
\beq
\ln \det AB=\ln \det A+\ln \det B+a_z(A,B)
\:,\label{e10}\eeq
or
\beq
a_z(A,B)=-\ze'(0|AB)+\ze'(0|A)+\ze'(0|B)
\:.\label{e11}\eeq

Let us now show, in the simple example treated above, that other non-analytic 
regularizations 
are in fact affected by the same multiplicative anomaly.
Consider, to this aim, the so called proper-time regularization, one of the 
most widely used non-analytic regularizations. It is defined by
\beq
\ln \det A_\ep=-\int_\ep^\ii dt\,  t^{-1}\Tr e^{-tA}=-\int_0^\ii dt\,  
\theta(t-\ep)t^{-1}\Tr e^{-tA}
\:,\label{d1}\eeq
where the ultraviolet cutoff $\ep$ {\it cannot} be remove and controls the 
singularities of the integrand for small $t$. Let us compute the 
{\it finite part} 
defined by
\beq
a_\ep(A+V,A)=-\mbox{FP}\left\{ \ln \det [(A+V)A]_\ep- \ln \det (A+V)_\ep-\ln
 \det A_\ep 
\right\}
\:.\label{d2}\eeq
If this finite part is non zero, we prove the existence of a 
multiplicative anomaly 
within this regularization.
Making use of
\beq
e^{-ta}=\frac{1}{2\pi i }\int_{\Re z > 2}dz \Ga(z)t^{-z}a^{-z}
\:,\label{e12}\eeq
one easily obtains
\beq
a_\ep(A+V,A)=\mbox{FP}\left\{ \frac{1}{2\pi i }\int_{\Re z > 2}dz 
\frac{\Ga(z)}{z}\ep^{-z} \aq \ze(z|(A+V)A)-\ze(z|A+V)-\ze(z|A)\cq \right\}
\:.\label{e13}\eeq
Shifting the vertical contour to the left, one has simple poles at 
$z=2$ and  $z=1$ due to the zeta-functions, a double pole at $z=0$ and
 simple poles at 
$z=-1,-2,...$ due to $\Ga(z)/z$. The poles at $z=2$ , $z=1$ and $z=0$ 
give contributions proportional to  $\ep^{-2}$,  $\ep^{-1}$ and  $\ln \ep$,
representing the ultraviolet divergences, but the double pole at $z=0$ 
gives also a finite contribution. Thus, the  residues theorem  yields
\beq
a_\ep(A+V,A)=a_z(A+V,A)+\Ga'(1)\aq \ze(0|(A+V)A)-\ze(0|A+V)-\ze(0|A)\cq +O(\ep)
\:,\label{e14}\eeq
which shows that the multiplicative anomaly is present in this 
non-analytic regularization. The other finite contribution can be reabsorbed 
in the scale renormalization term $\ell^2$, since the partition function, 
in a more appropriate way, has to be written as \cite{hawk77-55-133}
\beq
\ln Z=-\frac{1}{2}\ln \det (A \ell^2)=\frac{1}{2}\ze'(0|A)-\frac{1}{2}\ze(0|A)\ln 
\ell^2
\:.\label{e34}\eeq
A similar argument is valid for other 
non-analytic regularizations (in the sense of changing the nature of the 
poles), as the dimensional regularization (see 
\cite{eliz97u-404}). 

As a consequence of this analysis, the question that one should answer 
is: can the multiplicative anomaly be reabsorbed into the 
renormalization procedure~? In a simple case (self-interacting scalar field 
in $\R^4$), since the multiplicative 
anomaly is a local functional, it does not contribute 
to the beta function at the one-loop approximation \cite{eliz97u-394}. 
However, in some case (as the relativistic ideal Bose gas), 
it may certainly have physical 
consequences, as shown in \cite{eliz97u-404} (see also \cite{adow1}).  

Our final conclusion is the following. Since it turns out that, as we have 
argued, there exist  potentially two different starting points, namely Eq.~(\ref{e1}) 
and Eq.~(\ref{e2}) 
as  formal definitions of the  (one-loop) partition function ---for the reasons explained 
above--- the dangerous aspects concerning the use of zeta-function 
regularization reported by Evans are 
not substantiated. Clear indications of this fact have been given in 
the present note. In our opinion, the zeta-function  method 
is a regularization procedure  on the 
same level as other regularization techniques within the definition of 
one-loop partition function given by Eq.~(\ref{e2}). It can be 
safely used ---provided one avoids  formal manipulations---
as has been done in the literature, with considerable successes, 
during the past twenty years.

\ack{We would like to thank G. Cognola, V. Moretti, R. Rivers and T. Evans 
 for valuable discussions. 
This work has been supported by the cooperative agreement
INFN (Italy)--DGICYT (Spain).
EE has been partly financed by DGICYT (Spain), project PB96-0925, and
by  CIRIT (Generalitat de Catalunya),  grant 1995SGR-00602. AF wishes 
to acknowledge financial support from the European Commission under TMR 
contract N. ERBFMBICT972020.
}

\end{document}